\documentclass[11pt]{llncs}
\usepackage{amssymb,amsmath,graphicx,color}
\usepackage{amsfonts}

\newcommand{\remove}[1]{}
\usepackage{epsfig,latexsym,graphicx}
\begin{document}

\title{Adjusting for Treatment Effects in Studies of Quantitative Traits}
\author{\textbf{Guide:} Prof. Saurabh Ghosh, Human Genetics Unit\\ \textbf{Project done by:} Subhabrata Majumdar, M.Stat. 2nd year}
\institute{M.Stat. final semester project\\Indian Statistical Institute, 203 B.T.Road, Kolkata - 700 108, India\\}
\maketitle

\begin{abstract}
A population-based study of a quantitative trait, e.g. Blood Pressure(BP) may be seriously compromised when the trait is subject to the effects of a treatment. Without appropriate corrections this can lead to considerable reduction of statistical power. Here we demonestrate this in the scenario of QTL mapping through Single-Marker Analysis. The data are simulated from a normal mixtrure for different values of allele frequencies, separation between normal populations and Linkage Disequilibrium, and several methods of correction are compared to check which can best compensate for the loss of power if treatment effects are ignored. In one of these methods, underlying BPs are approximated by subtracting an estimate of mean value of medicine effect from obsereved BPs in treated subjects. We domonestrate the efficacy of this method throughout different choices of parameters. Finally to account for quantitative traits that follow non-normal distributions, data are simulated from lognormal mixtures similarly and Kruskal-Wallis test is used to obtain estimates of powers for different methods of analysis.\\


\end{abstract}

\section{Introduction}
Quantitative traits are traits that take continuous values, like body height, body weight, Blood Pressure (BP) etc., and can be attributed to polygenic effects i.e. product of two or more genes. Population-based studies of a quantitative trait help to determine the genetic determinants of the trait. But in many such studies selective individuals receive treatments, like antihypertensive medicines for high BP. As a result, In treated subjects, the outcome of primary interest, the underlying BP,
which is the BP that an individual would have if he=she was not treated, cannot be measured and analysis must therefore be based on the observed BP. Without any correction, this leads to a reduction in statistical power and shrinkage in the estimated effects of determinants\cite{white}\cite{cui1}.

Some common and often-used ways to tackle this situation are as follows. One can completely ignore the information on treatment status and perform the analysis assuming that the observed values are same as the underlying values of the BP of the subjects\cite{brand}. Treated or affected subjects are sometimes excluded from the analysis\cite{rice}. Another common method is to adjust for BP treatment by incorporating it as a covariate \cite{odon}. One can also convert BP into a binary trait by defining as hypertensive the subjects who are treated or have an observed BP in excess of a stated threshold\cite{sethi}\cite{zhu}.

Apart from these, a general approach is \textit{Imputation}. Here the observed values are replaced by some plausible approximation of the actual BPs. The methods under this include addition of some statistic estimated from the data or some constant to the treated BP values \cite{cui2}, replacing the treated values by some random or constant quantity \cite{hunt}, adding residuals modified by a non-parametric algorithm \cite{levy}, and censored normal regression \cite{tobin}.

Here we first consider a simulation model for BP where observations are drawn from a normal mixture and are treated with medicine with a high probability if they cross a certain threshold. These are taken as observed BPs. Since in a practical scenario the genotype at the Quantitative Trait Locus (QTL) is not observed, genotypes of marker loci with different degrees of linkage with the QTL are considered and ANOVAs are calculated with respect to them. With this basic premise, different methods of analysis were applied on the simulated datasets and the resulting powers were compared. Lastly, we consider deviation from normality for underlying QT values. We now simulate from a lognormal mixture instead and then compare different methods of analysis through powers obtained through Kruskal-Wallis test, instead of ANOVA as before.

\section{Methods}
\subsection{The simulation model}
We consider our parameter of interest the systolic blood presure (SBP) of a subject. An individual is defined \textit{hypertensive} if the observed SBP exceeds a certain fixed threshold, which we take to be 140 mm Hg. In our case this ensures that the proportion of hypertensive individuals to the full population is not more than 0.2.\\

Suppose that the alleles of a gene affecting BP are $A$ and $a$, their frequencies being $1-p$ and $p$. Then the possible genotypes are $AA, Aa$ and $aa$. From Hardy Weinberg Equilibrium we know that these genotypes occur in the population with probabilities $(1-p)^2, 2p(1-p)$ and $p^2$, respectively. For the populations under these genotypes, we assume three equidistant normal distributions with means at $120 - d, 120$ and $120 + d$ mm Hg: $d$ being the distance between two adjacent means. With a given pair of values for $p$ and $d$, 1000 datasets each with 100 datapoints are simulated from the this normal mixture. Furthermore, a subject was declared hypertensive if the simulated BP exceeds 140 mm Hg. If deemed hypertensive, a subject was assigned to the treatment group with probability 0.8, and also a random treatment effect from a $N(-10,3^2)$ distribution is added to the underlying BP to obtain the observed BP. These simulations were performed with three choices of $p$ (0.1, 0.3, 0.5) and five choices of $d$ (10, 15, 20, 25, 30).

\subsection{Methods of analysis}
Seven different methods of analysis were compared on the simulated data sets. Except one method all others make use of the ANOVA model for calculating powers, with the effects of genotypes taken as fixed effects. The methods are described below. Consider in general $X_i, Y_i$ as the underlying and observed SBPs of the $i^{th}$ subject, respectively. We also define $M_i$ to be the indicator of the $i^{th}$ subject taking medicine.

\paragraph{\textbf{(a) Taking underlying BPs as observed:}}Although this is not feasible in practice, this is done to estimate the actual power of ANOVA on a sample from the full population.

\paragraph{\textbf{(b) No adjustment from treatment:}}We ignore the information on treatment status and perform the ANOVA assuming that observed BPs are same as underlying BPs, i.e. here $X_i = Y_i$ for all $i$. This gives an idea about the reduction of power due to treatment.

\paragraph{\textbf{(c) Omitting all affected individuals:}} Here we take $X_i = Y_i$ if $Y_i < 140$ and $M_i = 0$.

\paragraph{\textbf{(d) Omitting all treated individuals:}} We take $X_i = Y_i$ if $M_i = 0$.

\paragraph{\textbf{(e) Treatment effect modeled as a covariate:}}Here the underlying model is 
$$ X_i = \mu + \alpha_i + \beta M_i + \epsilon_i $$

and $X_i = Y_i$ for all $i$, with $\alpha_i$ being the fixed effect of genotype and $\beta$ being the random treatment effect. This is a mixed model, and the power is calculated considering the p-value of the F-statistic corresponding to the fixed effect.

\paragraph{\textbf{(f) Correction by a fixed quantity:}}The difference of means in treated and untreated but affected subjects, say $m$ is an unbiased and consistent estimate of the medicine effect (Proof in Appendix). We subtract this from each treated observation to get an estimate of the underlying BP and then perform the ANOVA, i.e. here we have $ X_i = Y_i - m M_i $.

\paragraph{\textbf{(g) Correction by a non-parametric algorithm:}} Here we use the non-parametric adjustment algorithm by Levy \textit{et al}\cite{levy} which has already been shown to give good approximations of actual powers in similar situation\cite{tobin}.

In this method, first observations are centered around the mean and raw residuals are obtained:
$$ r_i = Y_i - \bar{Y} $$

Then the residuals are ordered and modified residuals are obtained as follows (assume now that $\lbrace r_i\rbrace$ is the set of ordered residuals):
$$ r_k^* = (1 - M_i)r_k + M_i\left(\frac{r_k + \sum_{j=1}^{k-1} r_j^*}{k}\right) $$

The residuals are then sorted back to their original order and added to the mean observed BP to get estimates of the underlying BPs:
$$ X_i = \bar Y + r_i^* = Y_i - r_i + r_i^* $$

\subsection{Modifications for QTL mapping}

A caveat in the above approach is that in a practical situation, we do not exactly know the location of the Quantitative Trait Locus (QTL). Instead one can obtain genotypes at several marker loci near to the approximate position of the QTL within the genome. In that case, we perform the ANOVA with respect to each of these genotypes ignoring the effect of others and infer the QTL to be closest to the most significant marker within a given chromosomal region. This is called \textit{Single-Marker Analysis}. We now attempt to integrate this scenario into our approach.

Proximity of the marker loci to the QTL means a high degree of linkage among them, i.e. alleles at these two loci in an organism tend to pass on simultaneously to the next generation while reproducing. Now the effect of linkage between two loci in the genome is measured by a quantity called \textit{Linkage Disequilibrium} (LD). It measures the  of non-random association of alleles at two or more loci. Given the allele-pairs $A/a$ and $B/b$ the LD between two loci is defined as:
$$ \delta = P(AB) - P(A)P(B) $$

Where $P(AB)$ is the probability of the alleles $A$ and $B$ co-segregating, and $P(A), P(B)$ the respective individual allele frequencies. To make the value independent of the allele frequencies, $\delta$ is divided by its theoretical maximum to obtain a scale free quantity:
$$ \delta' = \begin{cases}
\frac{\delta}{\min\lbrace P(A)P(B),P(a)P(b)\rbrace} &\text{if }\delta < 0 \\
\frac{\delta}{\max\lbrace P(A)P(b),P(a)P(B)\rbrace} &\text{if }\delta > 0 \\
\end{cases}$$

Note that  $\delta'$ = 1 means complete linkage and $\delta'$ = 0 means no linkage at all i.e. independent assortment of alleles.

Our previous simulation model is now changed to incorporate this situation. Say we denote the alleles of the marker locus by $B/b$. Then under the above model, the frequencies of the four haplotypes are:
\begin{eqnarray*}
P(AB) &=& (1-p)^2 + \delta \\ P(Ab) &=& p(1-p) - \delta \\ P(aB) &=& p(1-p) - \delta \\ P(ab) &=& p^2 + \delta
\end{eqnarray*}
Now two biallelic haplotypes, each with a QTL and a marker allele, are generated and they are fused to obtain a biallelic genotype. Then the underlying BP value is generated from the distribution corresponding QTL genotype, but the observed genotype is taken as that of the marker locus. For example, if the haplotypes generated are $AB$, $aB$ and thus the genotype of a subject is $AaBB$, then the observation is taken from a $N(120 , 20^2)$ population which is the distribution corresponding to the QTL genotype $Aa$, but the observed genotype is taken as $BB$ and ANOVA is done based on the genotypes $BB, Bb$ and $bb$.\\

As before, 1000 datasets with 100 points each were simulated for $\delta'$ = 1/3 and 2/3. Note that our previous simulation corresponds to $\delta'$ = 1.

\section{Results and discussion}
All analyses were done on MATLAB version R2008a\cite{matlab}. For all the methods except (e), single-factor ANOVA was used to calculate the p-values, while for method (e), the p-value was calculated from the F-statistic of fixed effect in the mixed model. All significance levels were set at $\alpha = 0.05$. For a choice of $(p,d,\delta')$ the power to detect the effect of QTL, was therefore estimated as the proportion of datasets which were found to have the F-statistic with p-value $< 0.05$. The following three tables contain the powers for the three values of $\delta$ considered.

\begin{table}[!h]
\begin{center}
\begin{tabular}[c]{|c|c||c|c|c|c|c|c|c|}\hline
 & & \multicolumn{7}{c|}{Method of analysis (powers in percentage)}\\\cline{3-9}
\begin{large} $p$ \end{large} &\begin{large} $d$ \end{large} & All & All & Omit affected & Omit treated & Treatment & Constant & Non-parametric \\
 & (mm Hg) & underlying & observed & subjects & subjects & as covariate & adjustment & adjustment\\\hline\hline
 & 10 & 44.7 & 42.4 & 29.9 & 31.7 & 29.0 & 44.3 & 41.7\\\cline{2-9}
 & 15 & 81.2 & 79.1 & 60.8 & 68.6 & 59.5 & 80.9 & 79.8\\\cline{2-9}
\begin{large} \textbf{0.1} \end{large} & 20 & 97.3 & 97.0 & 86.1 & 89.9 & 85.9 & 97.3 & 97.1\\\cline{2-9}
 & 25 & 99.7 & 99.7 & 97.5 & 98.1 & 97.0 & 99.7 & 99.7\\\cline{2-9}
 & 30 & 100.0 & 100.0 & 99.7 & 99.7 & 99.5 & 100.0 & 100.0\\\cline{2-9}\hline\hline

 & 10 & 80.3 & 79.1 & 53.3 & 65.9 & 56.6 & 80.0 & 78.0\\\cline{2-9}
 & 15 & 98.9 & 97.8 & 90.2 & 93.9 & 91.6 & 98.6 & 98.8\\\cline{2-9}
\begin{large} \textbf{0.3} \end{large} & 20 & 100.0 & 100.0 & 99.3 & 99.8 & 99.7 & 100.0 & 100.0\\\cline{2-9}
 & 25 & 100.0 & 100.0 & 100.0 & 100.0 & 100.0 & 100.0 & 100.0\\\cline{2-9}
 & 30 & 100.0 & 100.0 & 100.0 & 100.0 & 100.0 & 100.0 & 100.0\\\cline{2-9}\hline\hline

 & 10 & 86.0 & 84.3 & 65.5 & 74.0 & 70.4 & 86.0 & 83.2\\\cline{2-9}
 & 15 & 99.7 & 98.8 & 95.2 & 97.6 & 96.9 & 99.7 & 99.5\\\cline{2-9}
\begin{large} \textbf{0.5} \end{large} & 20 & 100.0 & 100.0 & 99.9 & 100.0 & 100.0 & 100.0 & 100.0\\\cline{2-9}
 & 25 & 100.0 & 100.0 & 100.0 & 100.0 & 100.0 & 100.0 & 100.0\\\cline{2-9}
 & 30 & 100.0 & 100.0 & 100.0 & 100.0 & 100.0 & 100.0 & 100.0\\\cline{2-9}\hline
\end{tabular}
\caption{Powers obtained by different methods for $\delta' = 1$}
\end{center}
\end{table}

\begin{table}[!h]
\begin{center}
\begin{tabular}[c]{|c|c||c|c|c|c|c|c|c|}\hline
 & & \multicolumn{7}{c|}{Method of analysis (powers in percentage)}\\\cline{3-9}
\begin{large} $p$ \end{large} &\begin{large} $d$ \end{large} & All & All & Omit affected & Omit treated & Treatment & Constant & Non-parametric \\
 & (mm Hg) & underlying & observed & subjects & subjects & as covariate & adjustment & adjustment\\\hline\hline
 & 10 & 24.8 & 23.1 & 12.1 & 15.3 & 16.9 & 24.4 & 22.6\\\cline{2-9}
 & 15 & 45.5 & 43.1 & 24.3 & 32.0 & 27.9 & 45.3 & 41.7\\\cline{2-9}
\begin{large} \textbf{0.1} \end{large} & 20 & 68.3 & 64.8 & 42.1 & 49.5 & 46.5 & 67.2 & 64.5\\\cline{2-9}
 & 25 & 84.5 & 83.0 & 59.0 & 68.4 & 65.2 & 84.5 & 82.9\\\cline{2-9}
 & 30 & 92.7 & 92.4 & 74.7 & 78.4 & 80.7 & 92.5 & 92.0\\\cline{2-9}\hline\hline

 & 10 & 43.4 & 42.5 & 25.8 & 29.9 & 29.2 & 43.0 & 41.2\\\cline{2-9}
 & 15 & 75.1 & 74.9 & 47.7 & 58.1 & 56.0 & 75.1 & 73.8\\\cline{2-9}
\begin{large} \textbf{0.3} \end{large} & 20 & 92.9 & 92.8 & 71.7 & 81.1 & 76.9 & 92.6 & 91.8\\\cline{2-9}
 & 25 & 98.6 & 98.3 & 84.3 & 91.2 & 90.6 & 98.5 & 98.2\\\cline{2-9}
 & 30 & 99.5 & 99.5 & 93.3 & 96.3 & 94.9 & 99.5 & 99.5\\\cline{2-9}\hline\hline

 & 10 & 50.2 & 49.6 & 31.8 & 39.4 & 35.1 & 50.0 & 49.5\\\cline{2-9}
 & 15 & 82.1 & 80.9 & 58.3 & 68.7 & 61.4 & 82.1 & 79.2\\\cline{2-9}
\begin{large} \textbf{0.5} \end{large} & 20 & 95.9 & 95.6 & 79.4 & 89.0 & 82.8 & 95.7 & 94.3\\\cline{2-9}
 & 25 & 98.6 & 98.6 & 90.1 & 96.7 & 90.8 & 98.6 & 98.2\\\cline{2-9}
 & 30 & 99.8 & 99.8 & 93.6 & 98.1 & 96.1 & 99.9 & 99.6\\\cline{2-9}\hline
\end{tabular}
\caption{Powers obtained by different methods for $\delta' =2/3$}
\end{center}
\end{table}

\begin{table}[!h]
\begin{center}
\begin{tabular}[c]{|c|c||c|c|c|c|c|c|c|}\hline
 & & \multicolumn{7}{c|}{Method of analysis (powers in percentage)}\\\cline{3-9}
\begin{large} $p$ \end{large} &\begin{large} $d$ \end{large} & All & All & Omit affected & Omit treated & Treatment & Constant & Non-parametric \\
 & (mm Hg) & underlying & observed & subjects & subjects & as covariate & adjustment & adjustment\\\hline\hline
 & 10 & 8.9 & 7.9 & 6.8 & 7.7 & 6.5 & 8.5 & 7.3\\\cline{2-9}
 & 15 & 15.5 & 14.6 & 8.4 & 10.5 & 9.8 & 15.4 & 14.5\\\cline{2-9}
\begin{large} \textbf{0.1} \end{large} & 20 & 24.5 & 22.5 & 12.5 & 14.2 & 16.2 & 24.4 & 22.2\\\cline{2-9}
 & 25 & 32.1 & 30.4 & 17.2 & 20.5 & 19.6 & 32.0 & 30.0\\\cline{2-9}
 & 30 & 39.4 & 36.9 & 22.5 & 28.1 & 26.0 & 39.2 & 36.0\\\cline{2-9}\hline\hline

 & 10 & 12.9 & 12.5 & 9.9 & 10.3 & 8.5 & 12.2 & 12.5\\\cline{2-9}
 & 15 & 27.0 & 25.6 & 14.0 & 18.4 & 16.9 & 25.5 & 25.4\\\cline{2-9}
\begin{large} \textbf{0.3} \end{large} & 20 & 36.8 & 34.9 & 17.5 & 24.6 & 21.3 & 36.0 & 34.5\\\cline{2-9}
 & 25 & 46.8 & 45.7 & 24.3 & 29.8 & 28.9 & 46.3 & 43.9\\\cline{2-9}
 & 30 & 55.8 & 55.6 & 32.2 & 37.6 & 34.9 & 56.2 & 53.3\\\cline{2-9}\hline\hline

 & 10 & 14.7 & 15.0 & 11.1 & 13.5 & 12.7 & 14.4 & 14.8\\\cline{2-9}
 & 15 & 24.1 & 23.3 & 17.6 & 18.9 & 15.1 & 23.9 & 22.7\\\cline{2-9}
\begin{large} \textbf{0.5} \end{large} & 20 & 40.2 & 38.9 & 27.6 & 29.6 & 26.0 & 40.2 & 37.0\\\cline{2-9}
 & 25 & 50.6 & 50.2 & 29.2 & 40.3 & 31.6 & 49.8 & 47.9\\\cline{2-9}
 & 30 & 59.7 & 56.6 & 32.8 & 41.8 & 36.3 & 58.5 & 54.3\\\cline{2-9}\hline
\end{tabular}
\caption{Powers obtained by different methods for $\delta' = 1/3$}
\end{center}
\end{table}

First of all, as shown before\cite{col}, it is found that as the value of LD and thus degree of linkage between the QTL and marker locus decreases, the estimated power also falls for a given allele frequency and distance between marker genotype means. More importantly, from the above tables it is clear that there is reduction of power when single-marker ANOVA is performed considering the observed BPs instead of the true underlying BP values. Since a large proportion of hypertensive subjects, i.e. whose underlying BPs exceed the threshold of 140 mm Hg, are subjected to treatments that reduce the BP, a negative bias comes to the measurement of BP from these subjects. Without any adjustments this leads to shrinkage in the estimates of the effects of the genetic determinant\cite{tobin}.\\

Among the adjustment methods considered, the first two methods (coulmns 5 and 6 in the tables) are found to be very inefficient, understandably so because they selectively remove all or most of the affected individuals, who have high \textit{underlying} BP, thus discarding a lot of useful information.

The method of including treatment status as covariates also results in marked reduction of statistical powers. Although taking the information that whether a subject is being treated or not as a random effect in the assumed model seems to be a valid approach, it is actually flawed. This is so because a subject can receive treatment only if the underlying BP crosses a certain threshold, i.e. treatments are not assigned in a completely random way for all values of underlying BP. To be precise, if the underlying BP is $>$ 140 mm Hg, the probability of the subject receiving treatment is 0.8, but if it is less than the threshold, the probability is 0, i.e. none of these subjects are treated. Thus, rather than being an \textit{explanation} to the underlying BP, treatment status is an \textit{outcome} of it. Since much of the variability of the observed BPs is explained by underlying BPs, including treatment status as a covariate in a model with underlying BPs as the response variable is not a good idea.

Finally, correction by a fixed estimate of the effect of antihypertensive medicine gives the closest estimates of the true powers of the procedure in almost all the cases. The non-parametric adjustment algorithm by Levy \textit{et al}\cite{levy} also gives good estimates of the powers, but it is not much reliable since in most of the cases they are less than the powers when analysis is done ignoring treatment status.

\section{Deviation from normality}
Many quantitative traits follow non-normal distributions, like lognormal\cite{sil}, skew-normal\cite{fer} etc. For analyzing data on these traits from a population we cannot directly use ANOVA to obtain statistical power. To tackle this, one either has to go for ANOVA after some data transformation. In case a suitable  transformation is not found or there are issues regarding interpretation, non-parametric methods can be used to obtain estimates of powers.\\

For this situation, we simulate the datasets with the same choices of values for the parameters $(p,d,\delta')$, from a lognormal mixture population with the same values of means and variance as before. \textit{Kruskal-Wallis test} is used to obtain estimates of powers. Among the seven methods of analysis described before, all except the one considering treatment effect as covariate are used in this case as well. In addition, in the method of correcting by a fixed value, we obtain the estimate of medicine effect in a slightly different way. Instead of using the difference of mthe eans of observed BPs of treated and affected but untreated subjects, the difference of the corresponding medians is used. The results obtained are as follows.

\newpage\begin{table}[!h]
\begin{center}
\begin{tabular}[c]{|c|c||c|c|c|c|c|c|}\hline
 & & \multicolumn{6}{c|}{Method of analysis (powers in percentage)}\\\cline{3-8}
\begin{large} $p$ \end{large} &\begin{large} $d$ \end{large} & All & All & Omit affected & Omit treated & Constant & Non-parametric \\
 & (mm Hg) & underlying & observed & subjects & subjects & adjustment & adjustment\\\hline\hline
 & 10 & 43.3 & 42.2 & 35.8 & 36.6 & 43.3 & 43.3\\\cline{2-8}
 & 15 & 79.3 & 78.6 & 72.7 & 73.1 & 79.3 & 77.5\\\cline{2-8}
\begin{large} \textbf{0.1} \end{large} & 20 & 96.8 & 95.9 & 91.5 & 93.5 & 96.8 & 96.8\\\cline{2-8}
 & 25 & 99.6 & 99.1 & 98.9 & 99.8 & 99.6 & 99.6\\\cline{2-8}
 & 30 & 100.0 & 100.0 & 99.9 & 100.0 & 100.0 & 100.0\\\cline{2-8}\hline\hline

 & 10 & 79.8 & 79.0 & 69.6 & 72.5 & 79.8 & 78.5\\\cline{2-8}
 & 15 & 98.8 & 98.2 & 96.0 & 97.4 & 98.8 & 98.8\\\cline{2-8}
\begin{large} \textbf{0.3} \end{large} & 20 & 100.0 & 100.0 & 100.0 & 99.8 & 100.0 & 100.0\\\cline{2-8}
 & 25 & 100.0 & 100.0 & 100.0 & 100.0 & 100.0 & 100.0\\\cline{2-8}
 & 30 & 100.0 & 100.0 & 100.0 & 100.0 & 100.0 & 100.0\\\cline{2-8}\hline\hline

 & 10 & 85.5 & 85.4 & 76.8 & 82.3 & 85.5 &  83.3\\\cline{2-8}
 & 15 & 99.8 & 99.6 & 98.7 & 99.3 & 99.8 & 99.8\\\cline{2-8}
\begin{large} \textbf{0.5} \end{large} & 20 & 100.0 & 100.0 &  100.0 & 100.0 & 100.0 & 100.0\\\cline{2-8}
 & 25 & 100.0 & 100.0 &  100.0 & 100.0 & 100.0 & 100.0\\\cline{2-8}
 & 30 & 100.0 & 100.0 & 100.0 & 100.0 & 100.0 & 100.0\\\cline{2-8}\hline
\end{tabular}
\caption{Non-parametric case: powers obtained by different methods for $\delta' = 1$}
\end{center}
\end{table}

\begin{table}[!h]
\begin{center}
\begin{tabular}[c]{|c|c||c|c|c|c|c|c|}\hline
 & & \multicolumn{6}{c|}{Method of analysis (powers in percentage)}\\\cline{3-8}
\begin{large} $p$ \end{large} &\begin{large} $d$ \end{large} & All & All & Omit affected & Omit treated & Constant & Non-parametric \\
 & (mm Hg) & underlying & observed & subjects & subjects & adjustment & adjustment\\\hline\hline
 & 10 & 21.1 & 20.5 & 15.3 & 19.8 & 21.1 & 20.6\\\cline{2-8}
 & 15 & 44.8 & 44.5 & 33.8 & 35.2 & 44.5 & 44.1\\\cline{2-8}
\begin{large} \textbf{0.1} \end{large} & 20 & 63.5 & 63.0 & 52.5 & 58.4 & 63.5 & 63.2\\\cline{2-8}
 & 25 & 77.9 & 77.7 & 70.1 & 71.4 & 78.0 & 76.8\\\cline{2-8}
 & 30 & 90.0 & 89.9 & 79.2 & 79.4 & 89.7 & 89.6\\\cline{2-8}\hline\hline

 & 10 & 45.3 & 45.2 & 34.7 & 40.0 & 45.1 & 43.6\\\cline{2-8}
 & 15 & 75.7 & 75.0 & 63.0 & 67.0 & 75.0 & 72.8\\\cline{2-8}
\begin{large} \textbf{0.3} \end{large} & 20 & 94.3 & 94.2 & 82.1 & 85.1 & 94.3 & 93.5\\\cline{2-8}
 & 25 & 98.6 & 98.3 & 92.3 & 94.2 & 98.6 & 98.1\\\cline{2-8}
 & 30 & 99.9 & 99.7 & 96.8 & 97.7 & 99.8 & 99.5\\\cline{2-8}\hline\hline

 & 10 & 52.7 & 52.1 & 40.6 & 45.9 & 52.4 & 50.7\\\cline{2-8}
 & 15 & 84.3 & 83.0 & 69.6 & 77.4 & 84.3 & 81.2\\\cline{2-8}
\begin{large} \textbf{0.5} \end{large} & 20 & 95.5 & 95.1 & 88.0 & 91.0 & 95.3 & 94.3\\\cline{2-8}
 & 25 & 99.0 & 98.8 & 94.8 & 98.0 & 98.8 & 98.6\\\cline{2-8}
 & 30 & 99.9 & 99.8 & 97.0 & 98.8 & 99.8 & 99.6\\\cline{2-8}\hline
\end{tabular}
\caption{Non-parametric case: powers obtained by different methods for $\delta' = 2/3$}
\end{center}
\end{table}

\newpage\begin{table}[!h]
\begin{center}
\begin{tabular}[c]{|c|c||c|c|c|c|c|c|}\hline
 & & \multicolumn{6}{c|}{Method of analysis (powers in percentage)}\\\cline{3-8}
\begin{large} $p$ \end{large} &\begin{large} $d$ \end{large} & All & All & Omit affected & Omit treated & Constant & Non-parametric \\
 & (mm Hg) & underlying & observed & subjects & subjects & adjustment & adjustment\\\hline\hline
 & 10 & 8.2 & 7.7 & 6.4 & 6.3 & 8.3 & 7.2\\\cline{2-8}
 & 15 & 12.4 & 12.1 & 9.8 & 12.6 & 12.2 & 11.7\\\cline{2-8}
\begin{large} \textbf{0.1} \end{large} & 20 & 18.2 & 18.1 & 15.9 & 14.3 & 18.1 & 18.3\\\cline{2-8}
 & 25 & 27.8 & 27.1 & 16.9 & 18.0 & 27.6 & 27.0\\\cline{2-8}
 & 30 & 32.8 & 32.6 & 22.2 & 27.8 & 32.8 & 33.3\\\cline{2-8}\hline\hline

 & 10 & 12.7 & 12.1 & 9.8 & 11.8 & 12.7 & 12.1\\\cline{2-8}
 & 15 & 22.3 & 22.0 & 17.2 & 19.5 & 22.1 & 22.4\\\cline{2-8}
\begin{large} \textbf{0.3} \end{large} & 20 & 36.6 & 36.3 & 25.0 & 27.3 & 36.6 & 35.6\\\cline{2-8}
 & 25 & 39.7 & 39.7 & 33.0 & 35.0 & 39.7 & 38.4\\\cline{2-8}
 & 30 & 52.8 & 52.2 & 34.6 & 37.9 & 52.7 & 51.7\\\cline{2-8}\hline\hline

 & 10 & 13.5 & 12.8 & 13.1 & 13.7 & 13.5 & 13.4\\\cline{2-8}
 & 15 & 26.2 & 25.7 & 20.0 & 20.7 & 26.2 & 25.7\\\cline{2-8}
\begin{large} \textbf{0.5} \end{large} & 20 & 38.9 & 38.0 & 27.6 & 32.9 & 38.3 & 36.7\\\cline{2-8}
 & 25 & 50.1 & 49.5 & 36.8 & 40.5 & 49.7 & 47.8\\\cline{2-8}
 & 30 & 59.3 & 58.4 & 39.4 & 46.2 & 59.1 & 55.0\\\cline{2-8}\hline
\end{tabular}
\caption{Non-parametric case: powers obtained by different methods for $\delta' = 1/3$}
\end{center}
\end{table}

The powers obtained by different methods are comparable with the same powers when simulated from the normal mixture. In this case the reduction of power due to treatment effects seems to be less severe than the normal mixture simulation. The constant adjustment gives the closest powers to the estimated true powers, as before.

\section{Conclusion}
We have demonestrated through simulation that the distorting effect of antihypertensive therapy in studies of quantitatively measured blood pressure can lead to loss of statistical power in the Single-marker analysis approach of QTL mapping. Among the adjustment methods considered, ignoring the problem altogether and analysing observed BP in treated subjects as if it was the underlying BP, excluding affected or treated subjects from analysis, or fitting a mixed model with treatment as a binary covariate perform very poorly and thus should not be used. Finally we have concluded that adding an estimate of the reduced BP due to medicine can reasonably nullify the reduction of powers.

\section*{Acknowledgement}I thank my mentor, Prof. Saurabh Ghosh of Human Genetics Unit, Indian Statistical Institute, Kolkata for his unfailing help in guiding me through my doubts and providing ideas while doing this project.

\newpage\section*{Appendix: Properties of the estimate obtained in method (f) of section 2}

Suppose in a dataset there are total $n$ subjects, among whom $m$ are hypertensive and $k$ are treated with medicine. Given affected, a subject is treated with probability $p$, i.e. $k \thicksim Bin(m,p)$. Now, without loss of generality, say the subjects $1,...,m$ are affected, and among them $1,...,k$ are treated. $X_i, Y_i$ are the underlying and observed BPs of the $i^{th}$ subject, respectively.

For the sake of simplicity, we assume that underlying BP of all subjects come from a $N(\mu,\sigma^2)$ population, instead of the normal mixture considered in the main work. Subjects with underlying BP above a threshold $c$ are considered affected, and the medicine effects (say $B_i$, for the $i^{th}$ subject) are assumed to follow a $N(\nu,\tau^2)$ distribution. With this setting, our estimate under consideration will be:

$$\hat\nu = \frac{1}{k}\sum_{i=1}^k X_i - \frac{1}{m - k}\sum_{i=k+1}^m X_i $$

\paragraph{Unbiasedness:}$$$$
For $i = 1,...,k$ we have $X_i = Y_i + B_i$, and for $i = k + 1,...,m$, we have $X_i = Y_i$. Thus

$$ E(\hat\nu) = \frac{1}{k}\sum_{i=1}^k E(Y_i + B_i) - \frac{1}{m - k}\sum_{i=k+1}^m EY_i = \mu_c + \nu - \mu_c = \nu $$
with $\mu_c$ being the mean of the $N(\mu,\sigma^2)$ distribution truncated at $c$.

\paragraph{Consistency:}
\begin{eqnarray*}
Var(\hat\nu) &=& \frac{1}{k}\left(Var(Y_1) + Var(B_1)\right) + \frac{1}{m-k}Var(Y_1)\\
&=& \left(\frac{1}{k} + \frac{1}{m-k}\right)Var(Y_1) + \frac{1}{k}Var(B_1)\\
&=& \frac{m\sigma^2}{k(m-k)} + \frac{\tau^2}{k}\\
\Rightarrow \lim_{n\rightarrow\infty} Var(\hat\nu) &=& \lim_{m\rightarrow\infty}\frac{1}{m}\left[\frac{\sigma^2}{\frac{k}{m}\left(1-\frac{k}{m}\right)} + \frac{\tau^2}{\frac{k}{m}}\right]
\end{eqnarray*}
because as $n\rightarrow\infty, m\rightarrow\infty$.\\
Now $k\thicksim Bin(m,p)\Rightarrow \lim_{m\rightarrow\infty}\dfrac{k}{m} = p$, hence the quantity inside brackets will tend to $\sigma^2/p(1-p) + \tau^2/p$, i.e. a fixed quantity as $m\rightarrow\infty$. It follows that $\lim_{n\rightarrow\infty}Var(\hat\nu) = 0$.
\end{document}